\documentclass[aps,pra,reprint,amsmath,amssymb,superscriptaddress]{revtex4-2}

\usepackage{graphicx}
\usepackage{bm}
\usepackage{booktabs}
\usepackage{microtype}
\usepackage{xcolor}

\usepackage{soul}
\usepackage{xeCJK}
\usepackage{xcolor}
\usepackage[colorlinks,citecolor=blue,linkcolor=red,anchorcolor=blue,urlcolor=blue]{hyperref}

\newcommand{\kb}{k_{\mathrm B}}
\newcommand{\ee}{\mathrm e}
\newcommand{\dd}{\mathrm d}
\newcommand{\peq}{p_{\rm eq}}
\newcommand{\pa}{p_{\rm a}}
\newcommand{\Astore}{\mathcal A}
\newcommand{\Fneq}{\mathcal F}
\newcommand{\PhiB}{\Phi}
\newcommand{\Jd}{J_{\rm d}}
\newcommand{\Pc}{\mathcal P_{\rm c}}
\newcommand{\Pd}{\mathcal P_{\rm d}}

\begin{document}
\title{Active-Matter Battery}

\author{Yun-Qian Lin (林蕴芊)}
\altaffiliation{These authors contributed equally to this work.}
\affiliation{School of Physics and Astronomy, Beijing Normal University, Beijing, 100875, China}
\affiliation{Key Laboratory of Multiscale Spin Physics (Ministry of Education), Beijing Normal University, Beijing 100875, China}

\author{Rui-Han Liu}
\altaffiliation{These authors contributed equally to this work.}
\affiliation{School of Physics and Astronomy, Beijing Normal University, Beijing, 100875, China}
\affiliation{Key Laboratory of Multiscale Spin Physics (Ministry of Education), Beijing Normal University, Beijing 100875, China}

\author{Yu-Han Ma (马宇翰)}
\email{yhma@bnu.edu.cn}
\affiliation{School of Physics and Astronomy, Beijing Normal University, Beijing, 100875, China}
\affiliation{Key Laboratory of Multiscale Spin Physics (Ministry of Education), Beijing Normal University, Beijing 100875, China}

\begin{abstract}
Harnessing active matter to power functional devices has attracted growing interest. We propose a minimal battery model for storing energy extracted from active matter and derive its charging and discharging characteristics analytically. When the energetic cost of sustaining
activity is included, the charging power-efficiency trade-off becomes loop-like and the efficiency attains a finite-time maximum, in sharp contrast to passive energy devices.
\end{abstract}

\maketitle

\textit{Introduction.---}
Active matter continuously converts local energy consumption into
autonomous motion, force, and nonequilibrium fluctuations, from
molecular motors and microorganisms to synthetic colloids, tissues, and
driven granular systems
\cite{Marchetti2013,Bechinger2016,FodorCates2021}.
Its intrinsically nonequilibrium character generally requires an
energetic description beyond effective-temperature mappings
\cite{Fodor2016,Nardini2017,Mandal2017,PietzonkaSeifert2018}.
Active fluctuations and stresses can be rectified into persistent
currents \cite{Tailleur2009,Hanggi2009,Reichhardt2017}, drive
microscopic gears and rotors
\cite{Angelani2009,Sokolov2010,DiLeonardo2010}, and power cyclic or
autonomous engines
\cite{Krishnamurthy2016,Zakine2017,Holubec2020,
Pietzonka2019,FodorCates2021}.
Their thermodynamic cost, efficiency, and finite-time control have also
been increasingly clarified
\cite{Datta2022,Malgaretti2022,Markovich2021,Davis2024,Wang2025}.
Active matter is therefore well established as a medium for
nonequilibrium energy conversion.

Most active-matter energy-conversion devices use activity and extract work simultaneously: the active source remains coupled while useful
motion or power is produced. Here we propose the concept of an
\emph{active-matter battery}, in which active matter serves as the
charger and the acquired energy is retained for later use. Its defining feature is the temporal separation
between energy acquisition and energy use. The stored energy need not
be electrical and may be mechanical, chemical, or, more
generally, nonequilibrium free energy. Stochastic thermodynamics
provides the natural framework for quantifying work and free-energy
changes in finite systems
\cite{Jarzynski1997,Crooks1999,Seifert2012,Parrondo2015}, while quantum battery theory provides well-established operational
concepts for extractable work, charging power, and
controlled discharge \cite{Alicki2013,Campaioli2017,Ferraro2018,Quach2020,
Mitchison2021,Campaioli2024}. The central questions are therefore how much extractable work activity can store, how it can be retained after the charger is removed, and what power-efficiency trade-offs govern charging and discharging.

\begin{figure}[t]
\centering
\includegraphics[width=1\linewidth]{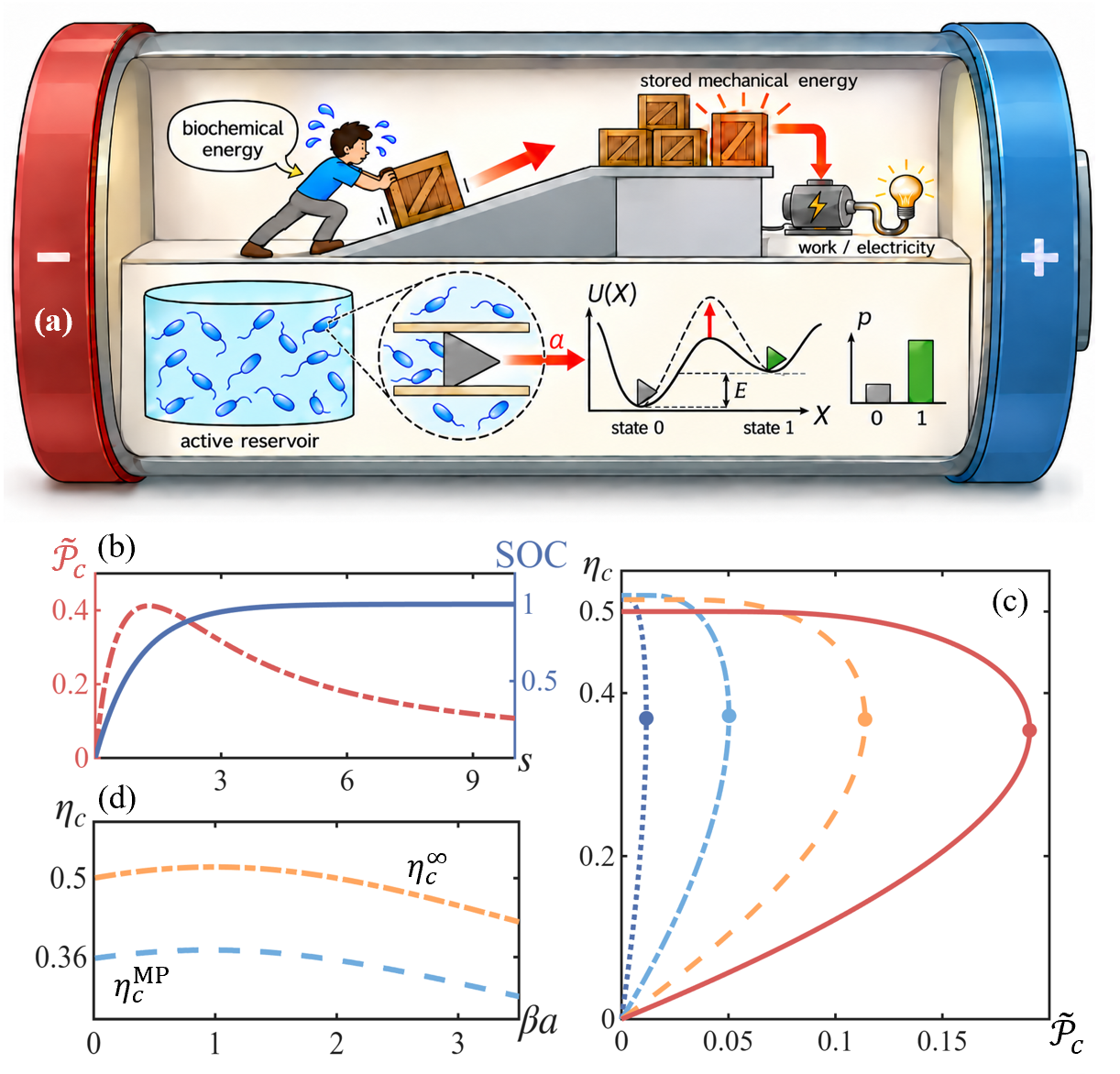}
\caption{(a) Schematic of the active-matter battery.
(b) Normalized state of charge and dimensionless charging power
$\tilde{\Pc}=\beta\Pc$ versus $s=\lambda_c t$ for $\beta a=4$.
(c) $\tilde{\Pc}$ versus charging efficiency $\eta_c$ for
$\beta a=0.5,1,1.5,$ and $2$, shown respectively by the blue dotted, blue dash-dotted, yellow dashed, and red solid curves. (d) Efficiency at maximum power $\eta_c^{\rm MP}$ and long-time
efficiency $\eta_c^\infty$ versus $\beta a$. Other parameters are fixed at $\beta E=1$.}
\label{fig:cycle}
\end{figure}

\textit{Minimal active-matter battery.---}We represent the storage element by a metastable two-state system with
energies $E_0=0$ and $E_1=E>0$. The upper-state occupation $p$ serves
as the charging coordinate, with thermal reference
$\peq=(1+\ee^{\beta E})^{-1}$ and $\beta=(\kb T)^{-1}$. As illustrated in Fig.~\ref{fig:cycle}(a), battery operation combines thermodynamic bias and kinetic gating. A nonequilibrium bias drives transitions toward the charged state; after the active source is removed, raising the barrier suppresses relaxation and retains the
stored resource; lowering the barrier while coupling to a load enables
discharge. These controls define the charge-store-discharge cycle.

For an occupation $p$, the nonequilibrium free energy is $\Fneq(p)=Ep+\kb T[p\ln p+(1-p)\ln(1-p)]$ \cite{Sekimoto1998,Jarzynski1997,Crooks1999,Esposito2010,
Esposito2011,Seifert2012,Parrondo2015}. We identify the stored work with the excess free energy above thermal equilibrium,
\begin{equation}
\Astore(p)
\equiv
\Fneq(p)-\Fneq(\peq)
=
\kb T D_{\rm KL}(P\Vert P_{\rm eq}),
\label{eq:storedWork}
\end{equation}
where $P=(p,1-p)$, $P_{\rm eq}=(\peq,1-\peq)$, and $D_{\rm KL}$ is
the Kullback-Leibler divergence
\cite{KullbackLeibler1951}. The maximum accessible $\Astore$ therefore sets the battery capacity. The variable conjugate to $p$ defines the generalized battery voltage,
\begin{equation}
\PhiB(p)
\equiv
\frac{\partial\Astore}{\partial p}
=E+\kb T\ln\frac{p}{1-p},
\label{eq:Phi}
\end{equation}
which vanishes at equilibrium, $\PhiB(\peq)=0$. Accordingly, $\Astore(p)=\int_{\peq}^{p}\PhiB(p')\dd p'$, so that $p$ and $\PhiB$ play the roles of generalized charge and voltage, respectively.

\textit{Charging with an active bath.---} Charging is achieved by an active affinity $a$, a generalized
nonequilibrium driving force \cite{DeDonder1936,Schnakenberg1976,Maes2021}. At the coarse-grained level, local detailed balance gives
\begin{equation}
\ln\frac{k_{01}^{\rm c}}{k_{10}^{\rm c}}
=
\beta(a-E),
\label{eq:chargeAffinity}
\end{equation}
so that $a$ is the work supplied by the active source per net
$0\to1$ transition. For symmetric rate splitting,
$k_{01}^{\rm c}=k_c\ee^{\beta(a-E)/2}$ and
$k_{10}^{\rm c}=k_c\ee^{-\beta(a-E)/2}$, where
$k_c=\sqrt{k_{01}^{\rm c}k_{10}^{\rm c}}$ is the kinetic rate scale.
For activated dynamics, $k_c\propto\ee^{-\beta B_c}$
\cite{Kramers1940,Hanggi1990}. Thus $a$ sets the thermodynamic bias,
whereas $B_c$ sets the charging time scale.

The charging dynamics follows the standard two-state Markov master
equation \cite{Esposito2010,Seifert2012},
$\dot p=k_{01}^{\rm c}(1-p)-k_{10}^{\rm c}p
=-\lambda_c(p-\pa)$.
Starting from the uncharged thermal state $p(0)=\peq$, its solution is
\begin{equation}
p(t)
=
\pa+(\peq-\pa)\ee^{-\lambda_c t},
\quad
\pa
=
\frac{1}{1+\ee^{\beta(E-a)}},
\label{eq:chargeDynamics}
\end{equation}
where $\lambda_c=k_{01}^{\rm c}+k_{10}^{\rm c}$.
Thus the affinity $a$ sets the stationary charged state $\pa$, whereas
the kinetic barrier $B_c$ controls the relaxation rate toward it. Population inversion occurs for $a>E$. Since
Eq.~\eqref{eq:Phi} gives $\PhiB(\pa)=a$, the affinity $a$ acts as the
open-circuit voltage imposed by the charger: charging stops when the
battery voltage reaches the charger affinity. To quantify the charging progress, we define the normalized state of charge (SOC) \cite{Plett2015},
\begin{equation}
y
\equiv
\frac{p-\peq}{\pa-\peq}
=
1-\ee^{-\lambda_c t},
\qquad
0\le y\le1,
\label{eq:SOC}
\end{equation}
which spans the accessible charging range from the thermal state $p=\peq$ ($y=0$) to the charger-defined state $p=\pa$ ($y=1$). Its evolution with the dimensionless time $s\equiv\lambda_c t$ is shown in Fig.~\ref{fig:cycle}(b).

The net charging current is
\begin{equation}
J_c(t)
=
k_{01}^{\rm c}[1-p(t)]
-k_{10}^{\rm c}p(t)
=
\dot p(t).
\label{eq:chargeCurrent}
\end{equation}
Each net $0\to1$ transition transfers the work $a$ from the active
source to the storage coordinate \cite{Esposito2009,Seifert2012}.
Hence the instantaneous input power delivered by the active charger to
the storage element is
$\dot W_c=aJ_c$, giving
\begin{equation}
W_c(t)
=
a[p(t)-\peq]
=
a(\pa-\peq)y.
\label{eq:chargeWork}
\end{equation}
Here $W_c$ is the work transmitted across the charger-battery
interface; it excludes the energetic cost of maintaining the active reservoir itself, which will be considered after the ideal
charge-store-discharge cycle.

We characterize the charging performance by the efficiency
$\eta_c=\Astore/W_c$ and the average stored-power rate
$\Pc=\Astore/t$, hereafter referred to as the charging power. Using Eq.~\eqref{eq:SOC},
\begin{equation}
\eta_c(y)
=
\frac{\Astore[p(y)]}{a(\pa-\peq)y},
\qquad
\Pc(y)
=
\lambda_c
\frac{\Astore[p(y)]}{-\ln(1-y)}.
\label{eq:chargeTradeoff}
\end{equation}
At short times, $p-\peq\propto t$ and
$\Astore\propto(p-\peq)^2$, so $\Pc\to0$; as $y\to1$, the stored free
energy saturates while the charging time diverges, again giving
$\Pc\to0$. Thus $\Pc$ peaks at a finite SOC, as shown by the red dash-dotted curve
in Fig.~\ref{fig:cycle}(b); the corresponding power-efficiency curves
for different affinities are shown in Fig.~\ref{fig:cycle}(c).

In the linear-response regime, $\beta a\ll1$, let
$q=\peq(1-\peq)$,
$\lambda_0=\lambda_c|_{a=0}$,
and $s=\lambda_0t$. Using
$\pa-\peq\simeq\beta q a$ and
$\Astore\simeq(p-\peq)^2/(2\beta q)$,
Eq.~\eqref{eq:chargeTradeoff} reduces to
\begin{equation}
\eta_c
=
\frac{1-\ee^{-s}}{2},
\qquad
\frac{\Pc}{\beta q a^2\lambda_0}
=
\frac{2\eta_c^2}{-\ln(1-2\eta_c)}.
\label{eq:chargeLR}
\end{equation}
The long-time efficiency ($s\rightarrow\infty$) approaches $\eta_c^\infty=1/2$ while the average charging power vanishes. Maximizing the charging power gives $\ee^{s_*}=1+2s_*$ with $s_*\simeq1.26$, yielding
\begin{equation}
\eta_c^{\rm MP}
\simeq
0.358,
\qquad
\Pc^{\max}
\simeq
0.204\,\beta q a^2\lambda_0.
\label{eq:chargeMPLR}
\end{equation}
The same power-efficiency relation and charging EMP arise in constant-voltage charging of an $RC$ circuit, where the stored energy is quadratic and the charging coordinate relaxes exponentially
\cite{Lei2025}. The value $\eta_c^{\rm MP}\simeq0.358$ therefore reflects this common single-exponential charging structure.

The finite-affinity curves in Fig.~\ref{fig:cycle}(c), however, show
that neither $\eta_c^{\rm MP}$ nor $\eta_c^\infty$ remains fixed at these linear-response values. As quantified in Fig.~\ref{fig:cycle}(d), both $\eta_c^\mathrm{MP}$ and $\eta_c^\infty$ vary nonmonotonically with $a$: they initially rise above $0.358$ and $1/2$, respectively, before decreasing at strong affinity as the charged occupation saturates. This finite-affinity behavior is analyzed in \textit{Appendix~A}.

\textit{Storage by kinetic locking.---} At a chosen charging time $t_c$, the active reservoir is removed at
$p_c=p(t_c)$, leaving the stored free energy $\Astore(p_c)$. Without further control, the charged occupation would relax toward $\peq$ and this resource would be lost. Storage therefore requires suppressing
interwell transitions while leaving the charged state unchanged. We accomplish this by raising the barrier from its charging value $B_c$ to a storage value $B_s$, while keeping the state energies $E_0$ and $E_1$ fixed. With the active affinity removed, the residual thermal rates satisfy detailed balance and can be written as $k_{01}^{\rm s}=k_s\ee^{-\beta E/2}$, $k_{10}^{\rm s}=k_s\ee^{\beta E/2}$,
$k_s\propto\ee^{-\beta B_s}$.
The corresponding leakage time is
$\tau_{\rm leak}=(k_{01}^{\rm s}+k_{10}^{\rm s})^{-1}
\propto\ee^{\beta B_s}$.
Thus increasing $B_s$ exponentially extends the retention time while
leaving the occupation $p_c$, and hence the stored free energy
$\Astore(p_c)$, unchanged. In this sense the charged occupation sets
the stored resource, whereas the kinetic barrier independently controls
its lifetime.

A useful locking protocol operates with barrier manipulation slow
compared with intrawell relaxation but fast compared with interwell
leakage. In this regime the occupation near the transition region
remains negligible while the barrier is raised. Importantly, kinetic
locking then need not require thermodynamic work on the storage system.
For a continuous potential $U(X,t)$, the work associated with barrier
manipulation is
$W_{\rm gate}=\int\dd t\int\dd X\,
\rho(X,t)\partial_tU(X,t)$.
If the potential is modified only near the barrier, where the
probability density is negligible, then $W_{\rm gate}\simeq0$.
Equivalently, in the coarse-grained Markov description, barrier
control changes transition rates but not state energies and therefore
performs no system work \cite{WangLu2026}. Related barrier-restoration operations are used to lock prepared states in bistable information memories \cite{Jun2014,Dago2023}. Here the
barrier instead preserves the nonequilibrium free energy deposited
during active charging. No active forcing or housekeeping power is
required during the hold stage. For a storage time
$t_{\rm h}\ll\tau_{\rm leak}$, the occupation remains
$p\simeq p_c$.

\begin{figure}[tbh!]
\centering
\includegraphics[width=1\linewidth]{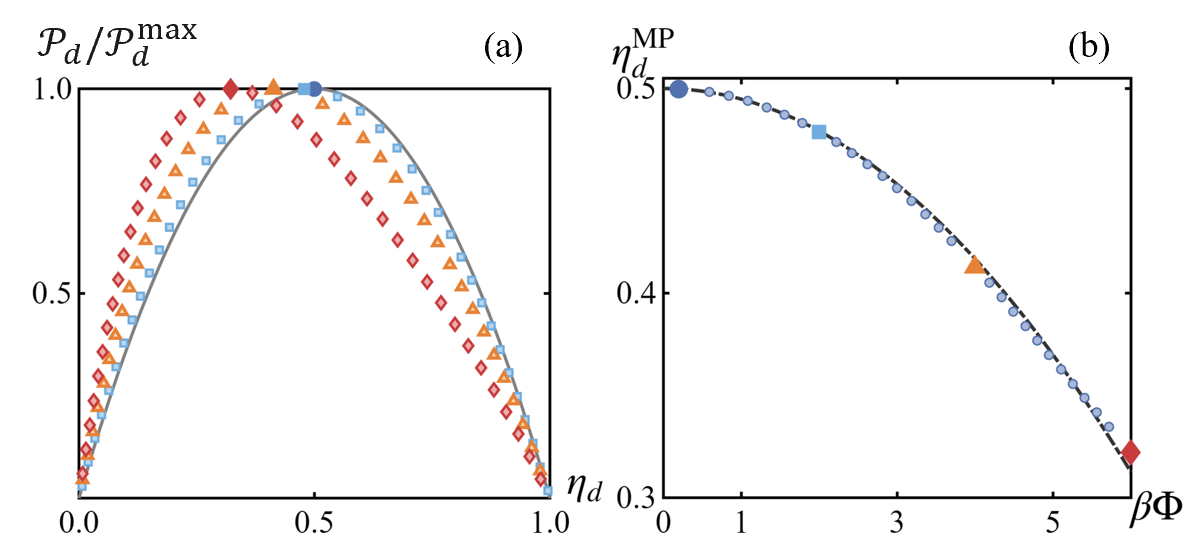}
\caption{(a) Normalized discharge power
$\mathcal{P}_d/\mathcal{P}_d^{\max}$ versus discharge efficiency $\eta_d=w/\PhiB$ for $\beta\PhiB=0.2$ (blue circles), $2.0$ (light-blue squares), $4.0$ (orange triangles), and $6.0$ (red diamonds). The gray solid curve denotes Eq.~\eqref{eq:approPower}. (b) Discharge efficiency at maximum power $\eta_d^{\rm MP}$ versus $\beta\PhiB$. Small dots show the exact numerical result, while the black dash-dotted curve shows $\eta_d^{\rm MP}=1/2-(\beta\PhiB)^2/192$. The large markers correspond to the maximum-power points of the trade-off curves in (a).}
\label{fig:performance}
\end{figure}

\textit{Discharging and powering an external load.---}Discharge reopens the kinetic channel by lowering the barrier from
$B_s$ to $B_d$ and coupling the storage coordinate to a load. Let
$w>0$ be the useful work delivered per net $1\to0$ transition.
Local detailed balance gives
$\ln(k_{10}^{\rm L}/k_{01}^{\rm L})=\beta(E-w)$.
For symmetric rate splitting,
$k_{10}^{\rm L}=k_d\ee^{\beta(E-w)/2}$ and
$k_{01}^{\rm L}=k_d\ee^{-\beta(E-w)/2}$, with
$k_d\propto\ee^{-\beta B_d}$. The discharge current is
$\Jd=k_{10}^{\rm L}p-k_{01}^{\rm L}(1-p)$. Using
Eq.~\eqref{eq:Phi}, $\beta\PhiB=\beta E+\ln[p/(1-p)]$, this becomes
\begin{equation}
\Jd
=
2k_d\sqrt{p(1-p)}
\sinh\!\left[
\frac{\beta}{2}(\PhiB-w)
\right].
\label{eq:current}
\end{equation}
Thus $\PhiB-w$ is the thermodynamic overpotential driving discharge.
For $0<w<\PhiB$ the battery delivers useful work, at $w=\PhiB$ the
current stalls, and for $w>\PhiB$ the transition is reversed. At fixed state of charge, the output power is $\Pd=w\Jd$ and the discharge efficiency is $\eta_d=w/\PhiB$. Defining
$x=\beta\PhiB/2$, we obtain
\begin{equation}
\Pd(\eta_d)
=
2k_d\sqrt{p(1-p)}\,\PhiB\,
\eta_d\sinh[x(1-\eta_d)].
\label{eq:exactPower}
\end{equation}
The power vanishes at $\eta_d=0$, where no useful work is delivered
per transition, and at $\eta_d=1$, where the current stalls. It
therefore reaches a maximum at an intermediate load, as shown in
Fig.~\ref{fig:performance}(a).

In the linear-response regime, $\beta\PhiB\ll1$,
Eq.~\eqref{eq:exactPower} reduces to
\begin{equation}
\Pd\simeq
\beta k_d\sqrt{p(1-p)}\,\PhiB^2\eta_d(1-\eta_d).
\label{eq:approPower}
\end{equation}
This symmetric parabolic relation $P\propto\eta(1-\eta)$ is analogous to that of finite-time heat engines near equilibrium
\cite{VandenBroeck2005,Shiraishi2016,RyabovHolubec2016,Ma2018Tradeoff,Qiu2025Roadmap} and also appears in conventional battery discharging \cite{Liu2026Discharging}, reflecting their common
linear force-current structure. It gives
$\eta_d^{\rm MP}=1/2$ and $\Pd^{\max}=\beta k_d\sqrt{p(1-p)}\PhiB^2/4$. This differs from the charging EMP $\eta_c^{\rm MP}\simeq0.358$, which arises from building a quadratic stored resource under exponential relaxation \cite{Lei2025}. As shown in Fig.~\ref{fig:performance}(a), for $\beta\PhiB=0.2$ the exact numerical result (blue squares) closely follows the linear-response parabola (gray line), validating Eq.~\eqref{eq:approPower}.

Beyond linear response, Eq.~\eqref{eq:current} retains the full nonlinear load-current relation. The trade-off curves in Fig.~\ref{fig:performance}(a) progressively depart from the symmetric parabola as $\beta\PhiB$ increases, with the maximum shifting to lower
efficiency. Writing $r=x\eta_d^{\rm MP}$, maximization of
Eq.~\eqref{eq:exactPower} gives
$x=r+\operatorname{artanh}r$ and
\begin{equation}
\eta_d^{\rm MP}
=
\frac{r}{r+\operatorname{artanh}r},
\qquad
\Pd^{\max}
=
\frac{4k_dr^2\sqrt{p(1-p)}}{\beta\sqrt{1-r^2}}.
\label{eq:EMPexact}
\end{equation}
Since $\operatorname{artanh}r>r$ for $0<r<1$,
$\eta_d^{\rm MP}<1/2$ at finite driving. Expanding in powers of
$\beta\PhiB$ gives
\begin{equation}
\eta_d^{\rm MP}
=
\frac12
-\frac{(\beta\PhiB)^2}{192}
-\frac{(\beta\PhiB)^4}{30720}
+O[(\beta\PhiB)^6].
\label{eq:EMPexpansion}
\end{equation}
Figure~\ref{fig:performance}(b) shows that the exact numerical EMP (blue dots) is accurately described near equilibrium by the quadratic approximation in Eq.~\eqref{eq:EMPexpansion} (black dash-dotted line), and decreases monotonically with increasing driving strength. The large
markers correspond to the values of $\beta\PhiB$ used for the trade-off
curves in Fig.~\ref{fig:performance}(a) .

Finally, the same battery voltage that determines the instantaneous
discharge current also bounds the total extractable work. During a
discharge from $p_c$ to $p_f$, spontaneous operation requires
$w(p)\leq\PhiB(p)$, and therefore
$W_{\rm out}=\int_{p_f}^{p_c}w(p)\dd p
\leq\Astore(p_c)-\Astore(p_f)$.
For complete discharge to equilibrium,
$W_{\rm out}^{\max}=\Astore(p_c)$. The bound is approached
quasistatically with $w(p)=\PhiB(p)-0^+$, for which the efficiency
tends to unity while the current and power vanish. 

\begin{figure}[tbh!]
\centering
\includegraphics[width=1\linewidth]{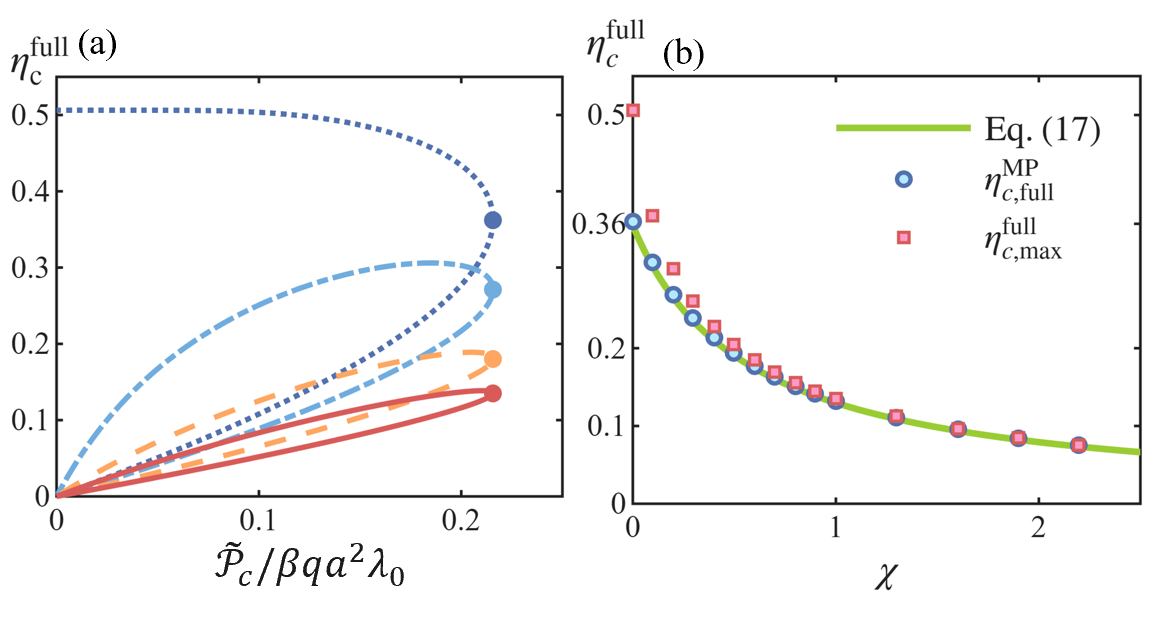}
\caption{(a) Dimensionless charging power
$\Pc/(\beta q a^2\lambda_0)$ versus full charging efficiency $\eta_c^{\rm full}$ for $\chi=0,0.2,0.6,$ and $1$ from top to bottom. (b) Efficiency at maximum power
$\eta_{c,\rm full}^{\rm MP}$ (blue circles) and maximum charging efficiency $\eta_{c,\max}^{\rm full}$ (red squares) versus $\chi$. The green solid curve shows Eq.~\eqref{eq:fullChargeEMP}. Other parameters are $\beta a=0.1$ and $\beta E=2$.}
\label{fig:housekeeping}
\end{figure}

\textit{Energetic cost of sustaining the active charger.---}
A physical active reservoir dissipates energy not only through useful
charging transitions, but also in maintaining its nonequilibrium
dynamics \cite{Davis2024,Wang2025}. We denote this background cost by
the housekeeping power $\mathcal P_{\rm hk}$. The corresponding
fuel-to-storage efficiency is
\begin{equation}
\eta_c^{\rm full}(t)
=
\frac{\Astore[p(t)]}
{a[p(t)-\peq]+\mathcal P_{\rm hk}t}.
\label{eq:fullChargeEff}
\end{equation}
Provided that $\mathcal P_{\rm hk}$ does not affect the charging
dynamics, it leaves $\Pc(t)$ unchanged but reshapes the
power-efficiency trade-off. As seen in Fig.~\ref{fig:housekeeping}(a), for $\mathcal P_{\rm hk}=0$(dark-blue dotted curve),
$\eta_c\to1/2$ as $\Pc\to0$, so the maximum efficiency is reached only
in the zero-power limit. For any $\mathcal P_{\rm hk}>0$, the stored free energy saturates while the housekeeping cost continues to grow, driving both $\Pc$ and $\eta_c^{\rm full}$ to zero at long times. The curve therefore closes into a loop. Increasing
$\mathcal P_{\rm hk}$ lowers the loop and the EMP, as seen from the curves from top to bottom, while leaving the maximum charging power unchanged.

To obtain explicit optimal-performance results, we specialize to linear response in the active affinity, $\beta a\ll1$, while keeping $\beta E$ arbitrary. To leading order, the EMP is
\begin{equation}
\eta_{c,\rm full}^{\rm MP}
\simeq
\frac{\eta_c^{\rm MP}}
{1+\kappa_*\chi},
\qquad
\kappa_*
=
\frac{s_*}{1-\ee^{-s_*}}
\simeq1.76,
\label{eq:fullChargeEMP}
\end{equation}
where $\chi\equiv\mathcal P_{\rm hk}/(\beta q a^2\lambda_0)$,
$q=\peq(1-\peq)$, and $\lambda_0=\lambda_c|_{a=0}$.
Thus $\eta_{c,\rm full}^{\rm MP}$ decreases from $\eta_c^{\rm MP}\simeq0.358$ at $\chi=0$ and scales as $0.204/\chi$ for strong housekeeping. As shown in Fig.~\ref{fig:housekeeping}(b), Eq.~\eqref{eq:fullChargeEMP} (green curve) agrees closely with the exact finite-affinity result at $\beta a=0.1$ (blue circles), confirming the linear-response prediction.

Housekeeping also creates a finite-time maximum efficiency. Unlike passive thermodynamic cycles or battery charging, where slower driving can approach the reversible limit \cite{Qiu2025Roadmap,Shiraishi2016,Lei2025}, here a longer charging time accumulates additional housekeeping loss. Within linear response,
the optimum $\eta_{c,\max}^{\rm full}=\eta_c^{\rm full}(s_\eta)$ satisfies $1-\ee^{-s_\eta}=\chi(\ee^{s_\eta}-1-2s_\eta)$ with $s_\eta>s_*$, so the maximum-efficiency point occurs later than the maximum-power point. For $\chi\ll1$, $\eta_{c,\max}^{\rm full}\to1/2$, whereas for $\chi\gg1$, $s_\eta\to s_*$ and both efficiencies approach $0.204/\chi$. The exact results in Fig.~\ref{fig:housekeeping}(b) show this crossover directly: increasing $\chi$ rapidly
brings $\eta_{c,\max}^{\rm full}$ (red squares) towards $\eta_{c,\max}^{\rm MP}$ (blue circles).

A derivation of Eq.~\eqref{eq:fullChargeEff} and the linear-response
results in Eq.~\eqref{eq:fullChargeEMP}, together with the condition
for $\eta_{c,\max}^{\rm full}$ and its limiting behaviors, is given in
\textit{Appendix~B}. The full-cycle fuel efficiency is correspondingly
lower than the storage-level efficiency because sustaining the active
charger incurs this additional housekeeping cost.

\textit{Microscopic realization of an active charger.---}
The theory above requires only the operational affinity $a$. A concrete
mechanical realization is obtained by taking the storage coordinate
$X$ to be spatial, with metastable states at $X_0$ and $X_1$ and
$\Delta X=X_1-X_0$. A homogeneous passive bath cannot produce a
directed force, since its uniform isotropic pressure satisfies
$\bm F_{\rm p}=-P\oint_{\partial\Omega}\bm n\,\dd S=0$.
By contrast, active pressure can depend on boundary interactions
\cite{Takatori2014,Solon2015,Solon2015NP}, allowing asymmetric
boundaries to rectify activity into mechanical work.

As illustrated in the lower panel of Fig.~\ref{fig:cycle}(a), we consider a triangular passive particle constrained to translate along
$X$ without rotating. Its asymmetric faces interact differently with the surrounding active particles, leading to unequal active pressures. To parameterize this effect, consider elliptical active Brownian particles near a soft boundary~\cite{Solon2015NP}, for which $P(\lambda)=P_0f(z)$ with $P_0=\rho v_0^2/(2\mu_tD_r)$, $f(z)=(1-\ee^{-z})/z$, and
$z=\lambda\mu_r\kappa/D_r$.. Here $\rho$ and $v_0$ are the particle density and propulsion speed, $\mu_{t,r}$ the translational and rotational mobilities, $D_r$ the rotational diffusivity, and $\kappa$ the particle anisotropy, while $\lambda$ characterizes the effective boundary coupling. Different couplings on the two faces then produce $\Delta P=P_0\mathcal{G}$, with $\mathcal{G}\equiv f(z_L)-f(z_R)$ the dimensionless boundary-asymmetry factor. The resulting directed force is $F_a=L_\perp\Delta P$, where $L_\perp$ is the transverse size of the
particle. 

If $F_a$ varies weakly between the two metastable positions, the affinity is the mechanical work performed by the active bath,
\begin{equation}
a
=
\int_{X_0}^{X_1}F_a(X)\dd X
\simeq
\frac{\rho v_0^2\mathcal{G}L_\perp\Delta X}
{2\mu_tD_r}.
\label{eq:microa}
\end{equation}
Equation~\eqref{eq:microa} directly connects the coarse-grained affinity to microscopic activity and boundary geometry: $a$ grows
with the active-pressure scale $\rho v_0^2/D_r$, the geometric
asymmetry $\mathcal{G}$, and the displacement scale
$L_\perp\Delta X$. For reference, the asymmetric partition considered in Ref.~\cite{Solon2015NP}, with $z_L=1$ and $z_R=4$, gives $\mathcal{G}\simeq0.39$, indicating that boundary asymmetry can produce a pressure contrast of order $P_0$. At high density, however, collisional slowing eventually reduces the active pressure \cite{Takatori2014,Solon2015}. A Kramers derivation of Eq.~\eqref{eq:microa} and its regime of validity are given in
\textit{Appendix~C}. This construction provides one microscopic realization of $a$; the general theory does not rely on active
pressure.

\textit{Concluding remarks.---}
We proposed a minimal active-matter battery that stores work extracted from active reservoir as metastable nonequilibrium free energy and releases it controllably to a load. The affinity sets the thermodynamic
driving and accessible stored resource, while the barrier controls the
kinetic time scales without changing the state energies:
\begin{center}
\begin{tabular*}{0.98\columnwidth}{@{\extracolsep{\fill}}ccc@{}}
\toprule
\textbf{Stage} & \textbf{Environment} & \textbf{Battery gate} \\
\midrule
Charge    & Affinity $a$ & $B_c$ \\
Storage   & $a=w=0$      & $B_s\gg B_c,B_d$ \\
Discharge & Load $w$     & $B_d$ \\
\bottomrule
\end{tabular*}
\end{center}

The model yields explicit charging and discharging performance. Near
equilibrium, $\eta_c^{\rm MP}\simeq0.358$ and
$\eta_d^{\rm MP}=1/2$, with the latter decreasing beyond linear
response. When the energetic cost of sustaining activity is included,
the charging power-efficiency relation becomes loop-like and the full
efficiency develops a finite-time maximum rather than a quasistatic
optimum. Kinetic locking exponentially extends the retention time without changing the stored free energy, and asymmetric active pressure
provides a microscopic realization of the charging affinity. 

Active work may also be stored in chemical, elastic, or collective many-body states beyond the mechanical realization considered here. Interactions may provide collective thermodynamic advantages \cite{Liang2025,Rolandi2023,Vroylandt2018}, enhancing
finite-time charging and potentially inducing collective charging transitions . Time-dependent control of the active drive, barrier, and load may further optimize charging speed and fuel consumption
\cite{Schmiedl2007,SivakCrooks2012,Davis2024,Wang2025}. These directions point toward many-body and optimally controlled active-energy storage.

\textit{Acknowledgment.---}We thank L.T. Chen for fruitful comments. This work is supported by the National Natural Science Foundation under Grant No. 12305037. The authors used ChatGPT (OpenAI) for language polishing and assistance with coding and debugging. The scientific question, conceptual framework, and theoretical development were independently formulated by the authors.
All scientific content, numerical methods, and interpretation of the
results were checked and finalized by the authors. The schematic in
Fig.~\ref{fig:cycle}(a) was designed by Y. H. Ma and rendered with assistance from ChatGPT.

\bibliography{references}

\clearpage
\onecolumngrid{}
\begin{center}
\textbf{End Matter}
\end{center}
\twocolumngrid{}
\textit{Appendix A: Charging efficiency beyond linear response.---}We briefly examine charging beyond linear response regime. Let
$\alpha=\beta a$, $\epsilon=\beta E$, $p_0=\peq$,
$q=p_0(1-p_0)$, and
$m=1-2p_0=\tanh(\epsilon/2)$. The affinity-biased stationary occupation is
$p_a=[1+\ee^{\epsilon-\alpha}]^{-1}$, while
$p(y)=p_0+(p_a-p_0)y$ with
$y=1-\ee^{-\lambda_c t}$. Since
$\beta\Astore=D(p\Vert p_0)$, the exact efficiency and charging power are
\begin{equation}
\eta_c(y)=
\frac{D[p(y)\Vert p_0]}{\alpha(p_a-p_0)y},
\qquad
\Pc(y)=\frac{\lambda_c}{\beta}
\frac{D[p(y)\Vert p_0]}{-\ln(1-y)}.
\label{eq:chargeExactApp}
\end{equation}

The prefactor $\lambda_c$ does not affect the state $p_*$ at maximum
power. Using
$\partial_pD=\beta\PhiB(p)$, the stationarity condition becomes
\begin{equation}
D(p_*\Vert p_0)
=
(p_a-p_*)\ln\!\frac{p_a-p_0}{p_a-p_*}\,
\beta\PhiB(p_*),
\label{eq:chargeMPexactApp}
\end{equation}
which determines the EMP 
\begin{equation}
\eta_c^{\rm MP}=\frac{D(p_*\Vert p_0)}
{\alpha(p_*-p_0)}.
\end{equation}
exactly at arbitrary affinity. To expose the departure from linear response, expanding
$p_a-p_0=q\alpha+qm\alpha^2/2+\cdots$ and
\[
D[p(y)\Vert p_0]
=
\frac12qy^2\alpha^2
+\frac16qm\,y^2(3-y)\alpha^3
+O(\alpha^4)
\]
and solving the maximum-power condition perturbatively gives
\begin{equation}
\eta_c^{\rm MP}
=
0.358
+
0.065\,
\tanh\!\left(\frac{\beta E}{2}\right)\beta a
+O[(\beta a)^2].
\label{eq:chargeMPexpApp}
\end{equation}
Here the zeroth-order optimum satisfies
$\ee^{s_0}=1+2s_0$, with $s_0\simeq1.25643$.
Thus, for $E>0$, the EMP initially increases above its
linear-response value $0.358$.

The corresponding long-time zero-power efficiency $\eta_c^\infty
=D(p_a\Vert p_0)/\alpha(p_a-p_0)$ follows by setting
$p=p_a$, which is specifically 
\begin{equation}
\eta_c^\infty=
\frac12
+\frac{1}{12}\tanh\!\left(\frac{\beta E}{2}\right)\beta a
-\frac{q}{12}(\beta a)^2
+O[(\beta a)^3].
\label{eq:chargeInfApp}
\end{equation}
It therefore likewise rises above the linear-response value $1/2$. At strong affinity, however, $p_a\to1$ and the stored free energy saturates while the input work continues to grow with $a$:
$\eta_c^\infty\sim
\ln(1/p_0)/[(1-p_0)\beta a]$ and
$\eta_c^{\rm MP}\propto(\beta a)^{-1}$.
Both efficiencies are consequently nonmonotonic in $a$, with maxima at finite affinity.

\textit{Appendix B: Charging efficiency with housekeeping dissipation.---}The charging analysis in the main text first treats the active
reservoir as an ideal source, so that its energetic cost is given by
the work transmitted to the storage coordinate. A physical active
reservoir also consumes fuel to sustain its nonequilibrium dynamics,
even when no useful charging transition occurs \cite{Davis2024}.
This produces a time-extensive housekeeping dissipation, distinct from
the finite-rate cost of changing the storage state
\cite{Wang2025}. We denote the corresponding housekeeping power by
$\mathcal P_{\rm hk}$.

We assume that $\mathcal P_{\rm hk}$ is approximately constant during
one charging stroke and does not modify the coarse-grained transition
rates. The charging trajectory therefore remains
$p(t)=\pa-\Delta p\,\ee^{-\lambda_ct}$, with
$\Delta p=\pa-\peq$. Using
$y=(p-\peq)/\Delta p=1-\ee^{-\lambda_ct}$, the total fuel input is
$W_c^{\rm full}=a\Delta p\,y+\mathcal P_{\rm hk}t$, and the
fuel-to-storage efficiency becomes
\begin{equation}
\eta_c^{\rm full}
=
\frac{\Astore[p(t)]}{W_c^{\rm full}}
=
\frac{\Pc}
{a\Delta p\,y/t+\mathcal P_{\rm hk}}.
\label{eq:app_full_def}
\end{equation}
Here $\Pc=\Astore/t$ is the useful storage power, whereas
$\mathcal P_{\rm hk}$ is consumed without increasing the stored
resource. Using $\eta_c=\Astore/(a\Delta p\,y)$ gives
\begin{equation}
\eta_c^{\rm full}(y)
=
\frac{\eta_c(y)}
{1+\displaystyle
\frac{\mathcal P_{\rm hk}}
{a\Delta p\,\lambda_c}
\frac{-\ln(1-y)}{y}}.
\label{eq:app_exact_reduction}
\end{equation}
The housekeeping correction therefore grows with charging time:
the stored resource saturates, whereas the maintenance cost continues
to accumulate.

In the linear-response regime of $\beta a\ll1$. Defining
$q=\peq(1-\peq)$, $\lambda_0=\lambda_c|_{a=0}$, and
$s=\lambda_0t$, one has $\Delta p=\pa-\peq\simeq q\epsilon, p(t)-\peq\simeq
q\epsilon(1-\ee^{-s})$. Expanding the nonequilibrium free energy to quadratic order around
$\peq$ then gives $\Astore\simeq
\beta q a^2(1-\ee^{-s})^2/2$. Introducing
$\chi=\mathcal P_{\rm hk}/(\beta q a^2\lambda_0)$, the leading-order
charging power and full efficiency are
\begin{equation}
\frac{\Pc(s)}
{\beta q a^2\lambda_0}
=
\frac{(1-\ee^{-s})^2}{2s},
\eta_c^{\rm full}(s)
=
\frac{(1-\ee^{-s})^2}
{2[1-\ee^{-s}+\chi s]}.
\label{eq:app_LR_performance}
\end{equation}

The charging power is independent of $\chi$ under the present
no-feedback assumption. Its maximum remains at
$\ee^{s_*}=1+2s_*$, with $s_*\simeq1.26$, giving
\begin{equation}
\eta_{c,\rm full}^{\rm MP}
=
\frac{\eta_c^{\rm MP}}
{1+\kappa_*\chi},
\qquad
\kappa_*
=
\frac{s_*}{1-\ee^{-s_*}}
\simeq1.76.
\label{eq:app_EMP}
\end{equation}
This reproduces Eq.~\eqref{eq:fullChargeEMP} of the main text. For
$\chi\gg1$,
$\eta_{c,\rm full}^{\rm MP}\simeq0.204/\chi$.

Housekeeping dissipation also qualitatively changes the charging
efficiency. For $\chi=0$,
$\eta_c=(1-\ee^{-s})/2$ increases monotonically toward $1/2$ as the
charging power vanishes. For any $\chi>0$, however,
$\eta_c^{\rm full}$ vanishes at both short and long times and therefore
attains a genuine finite-time maximum. This contrasts with passive heat
engines, whose maximum efficiency is reached in the reversible
quasistatic limit
\cite{Qiu2025Roadmap,Shiraishi2016,RyabovHolubec2016,Ma2018Tradeoff},
and with passive charging without a time-extensive maintenance cost
\cite{Lei2025}; related finite-duration optima also arise in
thermodynamic control of active matter \cite{Davis2024,Wang2025}.

Differentiating Eq.~\eqref{eq:app_LR_performance} gives
\begin{equation}
1-\ee^{-s_\eta}
=
\chi\left(\ee^{s_\eta}-1-2s_\eta\right),
\label{eq:app_etaopt}
\end{equation}
where
$\eta_{c,\max}^{\rm full}\equiv\eta_c^{\rm full}(s_\eta)$.
Since $\ee^s-1-2s<0$ for $0<s<s_*$, one has $s_\eta>s_*$.
Thus the maximum-efficiency point occurs later than the maximum-power
point, with
$\eta_{c,\max}^{\rm full}>\eta_{c,\rm full}^{\rm MP}$ in general.


For weak housekeeping, $\chi\ll1$,
\begin{equation}
s_\eta
\simeq
\ln\frac{1}{\chi},
\qquad
\eta_{c,\max}^{\rm full}
\simeq
\frac12
\left[
1-\chi\left(\ln\frac{1}{\chi}+1\right)
\right].
\label{eq:app_eta_weak}
\end{equation}
Hence
$\eta_{c,\max}^{\rm full}\to1/2$, whereas
$\eta_{c,\rm full}^{\rm MP}\to\eta_c^{\rm MP}\simeq0.358$ as
$\chi\to0$; the two optima therefore remain distinct. For strong housekeeping, $\chi\gg1$,
\begin{equation}
s_\eta
=
s_*+
\frac{1-\ee^{-s_*}}
{(2s_*-1)\chi}
+O(\chi^{-2}),
\quad
\eta_{c,\max}^{\rm full}
\simeq
\frac{0.204}{\chi}.
\label{eq:app_eta_strong}
\end{equation}
Thus $s_\eta\to s_*^+$, and the maximum-efficiency and maximum-power
operating points coincide asymptotically to leading order.

The constant-$\mathcal P_{\rm hk}$ approximation is the minimal model
needed to expose this effect. For a time-dependent maintenance cost,
one replaces $\mathcal P_{\rm hk}t$ by
$\int_0^t\mathcal P_{\rm hk}(t')\dd t'$ in
Eq.~\eqref{eq:app_full_def}. The distinction between transmitted work
and housekeeping dissipation remains valid provided that the affinity
$a$ accounts only for the work bias associated with net storage
transitions.

\textit{Appendix C: From microscopic active force to the charging affinity.---}Here we justify the coarse-grained affinity in Eq.~\eqref{eq:chargeAffinity} from an underlying mechanical active
force. Let $U_0(X)$ be the bare bistable potential of the storage
coordinate, with minima $X_0$ and $X_1$ satisfying
$U_0(X_1)-U_0(X_0)=E$. After eliminating the active bath, suppose that
the storage coordinate experiences a quasistatic mean active force
$F_{\rm a}(X)$ in addition to thermal fluctuations, $\gamma_X\dot X=-\partial_XU_0(X)+F_{\rm a}(X)+\xi_T(t)$, where
$\langle\xi_T(t)\xi_T(t')\rangle
=2\gamma_X k_BT\,\delta(t-t')$.
This description applies when the active boundary layer relaxes rapidly
compared with the storage dynamics and active force fluctuations do not
dominate the rare barrier-crossing events.

Along the one-dimensional storage coordinate, the mean active force
defines the effective potential
\begin{equation}
U_{\rm eff}(X)
=
U_0(X)-\int^X F_{\rm a}(X')\,\dd X'.
\end{equation}
For a saddle at $X^\ddagger$, Kramers theory gives \cite{Kramers1940}, to exponential
accuracy,
\begin{align}
k_{01}^{\rm c}
&\propto
\exp\!\left\{
-\beta\left[
U_{\rm eff}(X^\ddagger)-U_{\rm eff}(X_0)
\right]
\right\},
\nonumber\\
k_{10}^{\rm c}
&\propto
\exp\!\left\{
-\beta\left[
U_{\rm eff}(X^\ddagger)-U_{\rm eff}(X_1)
\right]
\right\}.
\end{align}
Taking their ratio eliminates the saddle energy and yields
\begin{equation}
\ln\frac{k_{01}^{\rm c}}{k_{10}^{\rm c}}
=
\beta\left[
\int_{X_0}^{X_1}F_{\rm a}(X)\,\dd X-E
\right],
\label{eq:rateRatioB}
\end{equation}
up to the usual prefactor corrections. Comparison with
Eq.~\eqref{eq:chargeAffinity} identifies $a=\int_{X_0}^{X_1}F_{\rm a}(X)\,\dd X$. Thus the charging affinity is the mechanical work performed by the coarse-grained active force during one forward charging transition.

For the pressure-driven realization discussed in the main text,
$F_{\rm a}(X)=L_\perp\Delta P(X)$, and therefore
$a=L_\perp\int_{X_0}^{X_1}\Delta P(X)\,\dd X$.
If the pressure imbalance is approximately constant,
\begin{equation}
a
\simeq
L_\perp\Delta X\,P_0
\left[f(z_L)-f(z_R)\right],
\qquad
P_0=\frac{\rho v_0^2}{2\mu_tD_r},
\label{eq:explicitAffinityB}
\end{equation}
which directly relates the coarse-grained affinity to the active
density, propulsion speed, persistence, and boundary asymmetry.

The reduction above also clarifies its range of validity. It requires
a separation between the relaxation time of the active boundary layer and the rare interwell transition time, so that $F_{\rm a}(X)$ is
well defined quasistatically. It further assumes that barrier crossing
is controlled predominantly by the thermal bath, while activity enters
mainly through the mean force. If colored active fluctuations or bath
memory contribute substantially to barrier crossing, the simple
Kramers form and $a=\int_{X_0}^{X_1}F_{\rm a}(X)\,\dd X$ need not hold.
The operational definition
$a=E+\kb T\ln(k_{01}^{\rm c}/k_{10}^{\rm c})$
from Eq.~\eqref{eq:chargeAffinity}, however, remains valid at the
coarse-grained level.
\end{document}